\documentstyle[12pt]{article}
\begin{document}
\newcommand{\be}{\begin{equation}}
\newcommand{\ee}{\end{equation}}
\mbox{ }\hfill{\normalsize ITP-93-5E} \\
\mbox{ }\hfill{\normalsize hep-th/9307184}
\mbox{ }\hfill{\normalsize January 1993} \\
\thispagestyle{empty}

\begin{center}
{\Large \bf Quantum Mechanics and Thermodynamics \\ of
Particles with Distance Dependent Statistics} \\
\vspace{1cm}
{\large Stefan V.~Mashkevich}\\[.5cm]
{\large \it
N.N.Bogolyubov Institute for Theoretical Physics, \\ 252143 Kiev,
Ukraine}
\end{center}
\vspace{.4cm}
\begin{abstract}
The general notion of distance dependent  statistics  in  anyon-like
systems is discussed. The two-body problem for  such  statistics  is
considered, the general formula for the second virial coefficient is
derived and it is shown that in the limiting cases it reproduces the
known results for ideal anyons.
\end{abstract}
\newpage

In recent years, essential progress had been achieved in solving
various problems of anyon quantum mechanics. Such problems arise
in two space dimensions and consist in finding those solutions of
a Schr\"odinger equation
\be
{\cal H}\Psi = E\Psi
\ee
which satisfy the interchange conditions
\be
{\cal P}_{jk}\Psi = \exp(i\pi\delta)\Psi
\label{Eq2}
\ee
for each  $j,k$ , where  ${\cal P}_{jk}$ is the operator of
anticlockwise  interchange of particles  $j,k \: .$ Up to now,  many
exact  and  approximate results have been obtained concerning the
two-anyon  \cite{1,2},  three-anyon \cite{3,4,5},
four-anyon \cite{9}, $N$-anyon \cite{10,11} problems,  and
connection with fractional quantum Hall effect \cite{12}
and high-$T_{c}$  superconductivity \cite{13} was widely
discussed.

It is, however, natural to put a question: Is the whole matter
relevant to real life? Indeed, there is no  doubt  that
bosons  and fermions do exist in nature, but there is  as
well  no doubt  that particles with "intrinsic" anyonic
statistics do  not  exist  ---  at least because the
real world is three-dimensional. It is  only  possible to
"imitate" this statistics with interaction. One possibility
(we do not argue it to be the only one) is  the
Chern-Simons  gauge field model. If a conserved current  $j_{\mu}$
is coupled to such field so that
\begin{eqnarray}
{\cal L} = {\cal L}_{G} - j_{\mu}A^{\mu} \;\;\; , \label {Eq3} \\
{\cal L}_{G} = {1 \over 2} \alpha \epsilon^{\mu\nu\lambda}
  A_{\mu} \partial_{\nu} A_{\lambda} \;\;\; ,
\end{eqnarray}
then a single charge $j_{\mu} = e \delta^{0}_{\mu} \delta^{2}(\vec{r})$
gives rise to the vector  potential with the only non-vanishing component
\be
A_{\phi}(\vec{r}) = - \frac{\Delta}{er}
\label{Eq5}
\ee
where  $\Delta = \frac{e^2}{2\pi\alpha} \;\; , \;\; r = |\vec{r}|\; .$
This is  precisely  the  potential  that figures in the
Aharonov-Bohm effect  and  leads  to  the  additional
interchange phase   $\delta  =  \Delta  .$  In  three
dimensions
such picture is possible if the objects in question are
stretched  in  one direction, i.e. if they are, for example,
vortex-like excitations.

It has been shown that under certain conditions the Chern-Simons
term can be generated as part of gauge field effective action
\cite{14}. In this way, anyons can arise
in real  physical  models.  From this point of view, we would
like to investigate the case  when  the gauge field Lagrangian
is not just the Chern-Simons term but  rather the sum
\be
{\cal L}_{G} =
{\cal L}_{0} + {1 \over 2} \alpha \epsilon^{\mu\nu\lambda}
  A_{\mu} \partial_{\nu} A_{\lambda} \;\;\; .
\label{Eq6}
\ee
The most natural candidate for ${\cal L}_{0}$
is the Maxwell term $- {1 \over 4} F_{\mu\nu}F^{\mu\nu} \; ,$
which can either be present in the theory from the
very beginning or be generated as a quantum correction, as
mentioned  in \cite{15}.  There may, however, be other
terms --- like, for example, $(\partial_{\mu}A^{\mu})^{2}$ in
Higgs-like models \cite{16}.

Let there again be a static charge in the origin. Solving  the
field equations  corresponding  to  the  Lagrangian
(\ref{Eq3}),(\ref{Eq6})  yields  $A_{\mu}(\vec{r})$ ---
the potential produced by this charge. If ${\cal L}_{0}$
does not contain time and angle  $\phi$  explicitly,
which we will assume to be the
case, then this potential depends only on  $r \; .$  In
Lorentz  gauge, consequently, its radial component
vanishes. Its temporal component is responsible for
charge-charge interaction analogous to the  Coulomb one. It has
nothing to do with statistics, and  in  the  present
consideration we will not take it into account (assuming,
for example, the values of the charges to be sufficiently small).
There remains the angular component which is  of main  interest
for us. Without any loss of generality we can write it as
\be
A_{\phi} (r) = -\frac{\Delta(r)}{er} \;\;\; .
\label{Eq7}
\ee
The function $\Delta(r)$ possesses clear physical sense:
$\Phi(r) \equiv -\frac{2\pi\Delta(r)}{e} $
is the flux of the magnetic field created by the charge
through  the circle of radius  $r \; .$ Therefore in
realistic cases it should be continuous and
finite including the extreme values
\be
\Delta_{0} = \Delta(0) \;\;\; , \;\;
\Delta_{\infty} = \Delta(\infty) \;\;\; .
\label{Eq8}
\ee
For example, in Maxwell-Chern-Simons theory
\cite{15,17}  $\Delta(r) = \frac{e^2}{2\pi\alpha} \cdot
\left[1-\alpha r K_1(\alpha r) \right] $ so that
$\Delta_0 = 0$ and $\Delta_{\infty} =
\frac{e^2}{2\pi\alpha} \; .$

Consider now a quantum mechanical problem involving two charges.
According to the general rules, the Hamiltonian of their
relative motion is
\be
\tilde{\cal H}_2 = \frac{p_r^2}{m} +
\frac{\left[ p_{\phi} + \Delta(r) \right]^2}{mr^2} +
V(\vec{r})
\label{Eq9}
\ee
where $\vec{r}$  is the relative radius vector and
$V(\vec{r})$  is the mechanical interaction potential.
If $V$ corresponds to a central force, the angular part
of the relative wavefunction is separated as
\be
\tilde{\Psi} = \chi \exp(i\ell\phi)
\ee
and the levels  $E^{\ell}_n$  are determined from the family
of  equations  of radial motion for partial waves
\be
\tilde{\cal H}^{\ell}_2 \chi = E^{\ell}_n \chi
\label{Eq11}
\ee
with
\be
\tilde{\cal H}^{\ell}_2 = \frac{p^2_r}{m} +
\frac{\left[ \ell + \Delta(r) \right]^2}{mr^2} + V(r) \;\;\; .
\label{Eq12}
\ee
Since the wave function behaves upon  interchange as in
(\ref{Eq2}) with $\delta = \ell \bmod 2 \;\; , \;\; \ell$
has to be even (odd) integer for bosons (fermions).

In the pure Chern-Simons case  $\Delta(r) \equiv \Delta =
\mbox{const} \; ,$ the  potential (\ref{Eq5}) is a pure gauge
and can be annihilated by a gauge  transformation. The
transformed Hamiltonian is obtained from (\ref{Eq9}) by
crossing out $\Delta(r)$ --- so that for  $V = 0$  it
corresponds  to non-interacting particles. Under the same
transformation, the wave function  changes so that for
the new one Eq.(\ref{Eq2}) takes place with  $\delta = \Delta$
for  bosons or $\delta = 1+\Delta$  for fermions, and
Eq.(\ref{Eq11}) for  the radial part does not change.
We will use the term "ideal anyons" for the particles  whose
wave function satisfies (\ref{Eq2}), and "$\delta$-anyons"
for ideal anyons with specific $\delta .$ In this way, bosons
or fermions interacting  via  the potential (\ref{Eq5}) are
completely equivalent to non-interacting  $\Delta$-anyons
and $(\Delta+1)$-anyons, respectively.

Thus, constant  $\Delta$  determines the statistics of
the particles. By analogy, one might speak of the
general case (\ref{Eq7}),(\ref{Eq9})  as  of the
one of particles with {\em distance  dependent  statistics.}
Indeed,  the term itself should not be understood  too
literally.  In  the case under consideration, the problem
cannot be reduced to  the  one  of non-interacting particles
as it can be  for ideal  anyons.  If  the particles are
somehow put into such conditions  that  $r$  can  vary
through a range in which  $\Delta(r) \simeq \overline{\Delta} \; ,$
then  they will behave like $\overline{\Delta}$-anyons.
(From now on we will assume the particles themselves to be
bosons, unless otherwise specified.) For other range of  $r$  the
"effective statistics" is other, and in this sense  it  "depends  on
distance". But if $\Delta(r)$  varies considerably in the
actual range  of  $r$ --- which, generally speaking, is most
likely to be the case --- then the problem is not reduced
to the one of ideal anyons  and  requires special consideration.

For the $N$-particle problem there is nothing  principally  new.
The Hamiltonian is
\be
{\cal H}_N = \sum_{j=1}^{N}
\frac{\left( \vec{p}_j - e\vec{A}_j \right)^2}{2m}
+ V(\vec{r}_1 , \ldots , \vec{r}_N)
\label{Eq13}
\ee
where
\be
\vec{A}_j = \sum_{\stackrel{\scriptstyle k=1}{k \neq j}}^{N}
\vec{A} (\vec{r}_j - \vec{r}_k ) \;\;\; ,
\label{Eq14}
\ee
$\vec{A} (\vec{r})$ being the vector potential produced
by a single charge  in the origin. For ideal anyons the
$\vec{A} (\vec{r}_j)$'s  can be eliminated  by  a  gauge
transformation, for distance depending statistics they cannot.

Generally speaking, a system characterized by the  Hamiltonian
(\ref{Eq13}) possesses three parameters of dimensionality
of length:  radius of the system  $R$  determined by the
function  $V \; ,$ its related average interparticle
distance  $\xi \sim R/ \sqrt{N} \sim n^{-1/2} \;\; , \; n$
being the number of particles per unit area, and
characteristic radius of interaction  $d$ such that
$\Delta(r)$  differs weakly from  $\Delta_0$ for  $r \ll d$
and from  $\Delta_{\infty}$ for  $r \gg d \; .$ For
example, in Maxwell-Chern-Simons theory $d = {1 \over \alpha} \; .$
The structure of the energy levels of the system depends on the
interrelation between these parameters. If  $d \gg R \; , $
the system behaves like that of  $\Delta_0$-anyons.
In the opposite limiting case $d \ll \xi \; , $  one
might expect to have  $\Delta_{\infty}$-anyons. The latter,
however, is in general not the case. Consider the problem
of two particles connected  by  a non-extensible thread,
i.e. the family of Hamiltonians (\ref{Eq11}) with
\be
V(r) = \left\{ \begin{array}{lcc}
        0 \;\; & , & \;\; r \leq R \\
        \infty \;\; & , & \;\; r > R
        \end{array} \right.  .
\label{eq15}
\ee
The interparticle distance is bounded above by  $R$  and
below by  the centrifugal barrier. For  $d \gg R$  all is
clear --- the particles "feel" only  $\Delta(r)$  with
$r \ll d \; ,$ that is,  $\Delta_0 \; .$ For $d \ll R$
the  situation is not so simple. The values of energy for
low-lying levels are of the order of $\frac{1}{mR^2} \; ,$
hence the ratio  $r \over R$   for  the corresponding  left
classical turning points is not much less than unity, so
$r \gg d \; .$ In such states the particles feel only
$\Delta(r) \simeq \Delta_{\infty}$ and therefore,  say,
at sufficiently low temperatures, indeed,  behave  like
$\Delta_{\infty}$-anyons. But for levels with energy
$\frac{K}{mr^2}$ and $\ell \sim 1$ the turning point is at
$r \sim \frac{R}{\sqrt{K}} \; ,$
and for sufficiently large $K$ it may be  $r \sim d$ or  even
$r \ll d \; .$ This means that at high temperatures the
particles may feel the whole range of  $\Delta(r)$  and as
a result not necessarily be  like $\Delta_{\infty}$-anyons.
Note also that if for some  $\ell \;\;\;\;  \Delta_0 = -\ell$
takes  place, the centrifugal barrier may be absent for this
$\ell$ (as it is absent  for bosons for  $\ell = 0$ ). This,
generally speaking, makes the properties of the particles
to be closer to those of bosons.

Thermodynamic properties of the system are influenced  by
one more parameter --- thermal wavelength
$\lambda = \sqrt{2\pi / mT} \; ;$ the behavior  of
the system depends on how large is $\lambda$ compared  to
other  distance scales. In this paper we will restrict
ourselves  to  the simplest case of high temperatures, when
$\lambda \ll \xi \; .$ In this case one can  apply the virial
expansion of the equation of state which we will write as
\be
\frac{p}{nT} = 1 + a_{2}\lambda^{2}n + a_{3}\lambda^{4}n^{2}
+ \ldots
\label{Eq16}
\ee
To calculate the second virial coefficient  $a_2$ it is
sufficient  to solve the two-particle problem. Using
$\tilde{Z}_{2} \left[ \Delta(r),R \right]$ to  denote
the two-particle relative partition function, one has
in accordance with the general theory \cite{18,2}
\be
a_{2} \left[ \Delta(r) \right] = \;
-{1 \over 4} \; - 2\lim_{R \rightarrow \infty}
\left\{ \tilde{Z}_{2} \left[ \Delta(r),R \right] -
\tilde{Z}^{Bose}_{2} \left[ R \right] \right\} \;\;\; .
\label{Eq17}
\ee
The exact expression for  $\tilde{Z}_2$ is
\be
\tilde{Z}_{2} \left[ \Delta(r),R \right] =
\sum_{\ell=-\infty}^{\infty} \tilde{Z}_{2}^{\ell}
\left[ \Delta(r),R \right]
\label{Eq18}
\ee
where the summation is over even (odd) values of $\ell$ for
bosons (fermions) and the partial wave contribution is
\be
\tilde{Z}_{2}^{\ell} \left[ \Delta(r),R \right] =
\sum_{n=0}^{\infty} \exp \left( -\beta E^{\ell}_{n} \right) \;\;\; ,
\label{Eq19}
\ee
$E^{\ell}_{n}$ being determined from Eq.(\ref{Eq11}).
There is no common way to calculate the levels, but since
we deal with high-temperature regime,  the semiclassical method
of Uhlenbeck and Beth \cite{19} may  be  applied  to the problem,
as it was pointed out  in \cite{20}.  In  this method  the
angular motion is quantized (summation over  partial waves  is not
replaced by integration) but the radial motion is considered
classically. That is, the exact value (\ref{Eq19}) is replaced by
\be
\tilde{Z}_{2}^{\ell} \left[ \Delta(r),R \right] =
\frac{1}{2\pi} \, \int_{-\infty}^{\infty} dp_r
\, \int_{0}^{R} dr \, \exp \left( -\beta \tilde{H}^{\ell}_{2}
\right) \;\;\; .
\label{Eq20}
\ee
Upon substituting (\ref{Eq12}) and integrating with
respect to  $p_r$ we obtain
\be
\tilde{Z}_{2}^{\ell} \left[ \Delta(r),R \right] =
\frac{1}{\sqrt{2}\lambda} \, \int_{0}^{R} \, dr \:
\exp \left\{ -\frac{\left[ \ell + \Delta(r) \right]^2}{2\pi}
\frac{\lambda^2}{r^2} \right\} \;\;\; .
\label{Eq21}
\ee
For  $\Delta(r) = \mbox{const}$ the $r$ integration can be performed
exactly. It is therefore convenient to rearrange the last equation
using the formula
\be
\int_{0}^{z}  f\left(x,v(x)\right) \: dx = F\left(z,v(z)\right) \: - \:
\int_{0}^{z}  G\left(x,v(x)\right) v'(x) \: dx
\label{Eq22}
\ee
where
\be
F(z,y) = \int_{0}^{z} f(x,y) \: dx \;\;\; , \;\;\;
G(x,y) = \frac{\partial F(x,y)}{\partial y} \;\; .
\label{Eq23}
\ee
Substituting the obtained expression in (\ref{Eq17}),
we come eventually to the result
\begin{eqnarray}
a_2 \left[ \Delta(r) \right] & = & -{1 \over 4} + \delta_{\infty}
- {1 \over 2}\delta_{\infty}^{2} \nonumber \\
& & - \sum_{\stackrel{\ell = -\infty}{even}}^{\infty} \;
\int_{0}^{\infty} \, \mbox{erfc}
\left[ \frac{\left| \ell + \Delta(r) \right|}{\sqrt{2\pi}}
\frac{\lambda}{r} \right] \: \frac{d}{dr} |\ell + \Delta(r)|
\: dr \;\;\; ,
\label{Eq24}
\end{eqnarray}
where $\delta_{\infty} = \Delta_{\infty} \bmod 2 \;\; , \;\;
\mbox{erfc}(z) = \frac{2}{\sqrt{\pi}} \: \int_{z}^{\infty} \:
\exp (-x^2) \: dx \;$ and the particles themselves are bosons.
The result for fermions is easily obtained by changing
$\Delta(r) \rightarrow  \Delta(r)+1 .$ Note that
Eq.(\ref{Eq24}) is invariant, as it should be, under the
changes  $\Delta(r) \rightarrow -\Delta(r)$ and
$\Delta(r) \rightarrow -\Delta(r) + 2 .$
Therefore we will assume that  $0 \le \Delta_0 < 2$
and  $\Delta_{\infty} > \Delta_0 .$

For  $\Delta(r) = \mbox{const}$  the last term vanishes,
and one comes  to the known exact quantum mechanical result
\cite{2}  for  the second  virial coefficient \cite{20}.
It is the last term that  is  characteristic  for
distance dependent statistics and is subject to analysis.
Its structure tells us that it depends essentially on the
interrelation  between  $\lambda$  and  $d .$ One may fix
$d$  and observe the behavior of this term with  $\lambda$
changing.

First, let  $\lambda \gg d.$ A considerable contribution
to the integral is given only by such values of  $r$
for which the argument of erfc is not much more than
unity. This can happen  (a) for  $r$
\raisebox{-1ex}{$ \stackrel{\textstyle > }{\sim } $} $ \lambda \:,$
(b) in the vicinity of the point
$r_0$ such that  $\Delta(r_{0}) = -\ell$  of width  of
the order of $\frac{r_0}{\lambda \left| \Delta '(r_0) \right|}$
and  (c) for  $0 \le r$ \raisebox{-1ex}
{$ \stackrel{\textstyle < }{\sim } $}
$ \rho$ if in this  region  $\left| \ell + \Delta (r) \right|$
behaves like  $kr\: .$ In the case  (a)  the  contribution  is
small because  $r \gg d$  and  $\frac{d}{dr} \Delta(r) \sim 0 .$
In the case (b) the width of the vicinity and consequently
its contribution tends to zero with  $\lambda$  increasing.
In the case (c) the contribution  can  be  evaluated  as
$k \rho \cdot \mbox{erfc} \left( \frac{k\lambda}{\sqrt{2\pi}}
\right) \; ,$ which again tends to zero as $\lambda \rightarrow
\infty \; .$ Thus,
\be
a_2 \left[ \Delta(r) \right] \;\; \raisebox{-1.57ex}{$ \stackrel{\textstyle
\longrightarrow}{\scriptstyle \lambda \rightarrow \infty} $} \;\;
-{1 \over 4} + \delta_{\infty} -{1 \over 2}\delta_{\infty}^2  \;\;\; ,
\label{Eq25}
\ee
i.e., in this limit we have   $\Delta_{\infty}$-anyons
(or,  which  is  the same, $\delta_{\infty}$-anyons).

Now, let  $\lambda \ll d \; .$ At first, for simplicity
we take  $\delta_{\infty} < 2  \;.$ In the integral with
$\ell = 0$  we replace erfc with unity, and the integral
becomes  $|\Delta_{\infty}| - |\Delta_0| = \delta_{\infty} - \delta_0
\; ,$ where  $\delta_0 = \Delta_0 \bmod 2 \;.$
The remaining sum can be evaluated as
\begin{eqnarray}
\sum_{\stackrel{\ell = -\infty}{even}}^{\infty}
\int_{0}^{\infty} \: \left\{ \mbox{erfc} \left[
\frac{\ell + \Delta(r)}{\sqrt{2\pi}} \frac{\lambda}{r} \right]
- \mbox{erfc} \left[\frac{\ell - \Delta(r)}{\sqrt{2\pi}}
\frac{\lambda}{r} \right] \right\} \: \frac{d}{dr} \Delta(r)
\: dr \nonumber \\
\simeq - \int_{0}^{\infty} \: dr \cdot {1 \over 2} \:
\int_{0}^{\infty} \: d\ell \cdot 2 \frac{2}{\sqrt{\pi}}
\exp \left[ -\frac{\ell^2 \lambda^2}{2\pi r^2}\right]
\frac{\Delta(r)}{\sqrt{2\pi}} \frac{\lambda}{r} \:
\frac{d}{dr} \Delta(r) \nonumber \\
= - \int_{0}^{\infty} \: \Delta(r) \: \frac{d}{dr}
\Delta(r) \: dr = {1 \over 2}\Delta_0^2 -
{1 \over 2}\Delta_{\infty}^2 = {1 \over 2}\delta_0^2 -
{1 \over 2}\delta_{\infty}^2 \;\;\; ,
\label{Eq26}
\end{eqnarray}
and substitution in (\ref{Eq24}) yields
\be
a_2 \left[ \Delta(r) \right] \;\;
\raisebox{-1.57ex}{$ \stackrel{\textstyle
\longrightarrow}{\scriptstyle \lambda \rightarrow 0} $} \;\;
- {1 \over 4} + \delta_0 -{1 \over 2}\delta_0^2 \;\;\; ,
\label{Eq27}
\ee
i.e., we have  $\delta_0$-anyons. For  $\Delta_{\infty} > 2$
the  derivation is somewhat more complicated, the terms with
$-\Delta_{\infty} < \ell \le 0$  having to be  considered
separately, but an accurate analysis \cite{21} shows that for
$(\Delta_{\infty} - \Delta_0)\lambda/d  \ll  1 $  the
result is still (\ref{Eq27}).

The results that we have obtained for the two limiting cases can
be understood using simple qualitative considerations. It  is  known
that at high temperatures the partition  function  of  the  Bose  or
Fermi gas in the first approximation coincides with that of the  gas
of classical particles interacting via an effective potential $v(r)$
which differs considerably from zero only for
$r$ \raisebox{-1ex}{$ \stackrel{\textstyle <}{\sim} $}
$\lambda$ \cite{18}. In the classical language, the
particles "feel themselves" bosons or fermions only when
approaching each other for a distance of
the order $\lambda \: .$ The same situation takes
place for ideal anyons; it is easy to  show
that the effective potential for them is given by
\be
v(r) = -T \ln \left[ 1 + (1-2\delta^2) \exp \left(
-\frac{2\pi r^2}{\lambda^2} \right) \right] \;\;\; .
\label{Eq28}
\ee
This means that in our case the  "effective  statistics"  should  be
$\Delta(r)$  averaged in a way over the range of the width
of the order  $\lambda$ but not  $\xi$  as one might
naively assume. Thus, our particles should behave, roughly
speaking, like  $\delta_{\rm eff}$-anyons where
$\delta_{\rm eff}$ is a quantity of the order of
$\left( \frac{1}{\lambda} \: \int_{0}^{\lambda} \: \Delta(r) \:
dr \right) \bmod 2 \; .$ This evidently  tends  to
$\delta_0$ for $\lambda \ll d$ and to $\delta_{\infty} $ for
$\lambda \gg d \; .$

In this paper we have developed the general notion  of  distance
dependent statistics, which can manifest itself, in  particular,  in
two-dimensional systems of particles coupled to a  gauge  field  the
Lagrangian of which contains, among others, the  Chern-Simons  term.
We have considered a gas of such particles at high temperatures  and
obtained the general formula for the second virial coefficient, from
which it is seen how the behaviour of the gas depends on temperature
and what are the limiting cases in which we effectively  have  ideal
anyons. A more detailed consideration will  be  presented  elsewhere
\cite{21}.

I would like to express my gratitude to  Prof.~G.M.Zinovjev  for
useful advice and support and to E.Gorbar for fruitful discussions.

\end{document}